\documentclass{article}

\PassOptionsToPackage{numbers}{natbib}

\usepackage[dblblindworkshop,final]{neurips_2025}
\workshoptitle{AI for Music}

\usepackage[utf8]{inputenc} 
\usepackage[T1]{fontenc}    
\usepackage{hyperref}       
\usepackage{url}            
\usepackage{booktabs}       
\usepackage{amsfonts}       
\usepackage{nicefrac}       
\usepackage{microtype}      
\usepackage{xcolor}         
\usepackage{multirow}
\usepackage{tabularx} 

\title{Persian Musical Instruments Classification Using Polyphonic Data Augmentation}

\author{%
  \begin{tabular}[t]{c}
    Diba Hadi Esfangereh$^{*}$, Mohammad Hossein Sameti$^{*}$, Sepehr Harfi Moridani,\\
    Leili Javidpour, Mahdieh Soleymani Baghshah\\[2pt]
    \textnormal{Department of Computer Engineering, Sharif University of Technology, Tehran, Iran}\\[3pt]
    \textnormal{\footnotesize $^{*}$Equal contribution}
  \end{tabular}
}

\usepackage{silence}
\begin{document}

\maketitle

\begin{abstract}
Musical instrument classification is essential for music information retrieval (MIR) and generative music systems. However, research on non-Western traditions, particularly Persian music, remains limited. We address this gap by introducing a new dataset of isolated recordings covering seven traditional Persian instruments, two common but originally non-Persian instruments (i.e., violin, piano), and vocals. We propose a culturally informed data augmentation strategy that generates realistic polyphonic mixtures from monophonic samples. Using the MERT model (Music undERstanding with large-scale self-supervised Training) with a classification head, we evaluate our approach with out-of-distribution data which was obtained by manually labeling segments of traditional songs. On real-world polyphonic Persian music, the proposed method yielded the best ROC-AUC (0.795), highlighting complementary benefits of tonal and temporal coherence. These results demonstrate the effectiveness of culturally grounded augmentation for robust Persian instrument recognition and provide a foundation for culturally inclusive MIR and diverse music generation systems. Code is available at: \hyperlink{https://github.com/dibahadie/Persian-Musical-Instruments-Classification-Using-Polyphonic-Data-Augmentation}{Github}
\end{abstract}

\section{Introduction}

Musical instrument classification is a core task in music information retrieval (MIR), underpinning applications such as automatic transcription, recommendation, audio analysis, and music generation, where understanding instrument timbres enables models to synthesize realistic outputs and have better captions for training text-to-music models \citep{Jamshidi2024,ChenLerch2022,Dubey2023,Gardner2024}. Despite significant progress in Western music (e.g., classification of violin, piano, guitar via CNN/RNNs), non-Western traditions, particularly modal systems like Persian Dastgāh, remain underrepresented in MIR research \citep{musicalinstrumentidentification, Taenzer2023,Moysis2023,Abbasi2014,Azar2018,Farhat1990}. Persian music features microtonal nuances, ornamentation, and heterophonic textures that pose unique challenges for instrument identification.
Recent works begin to address these gaps. The Persian Piano Corpus (PPC) provides a Dastgāh-annotated dataset with instrument-level metadata, expanding resources available for modal and instrument studies \citep{persianpianocorpus}. 
Nevertheless, most instrument classification models remain monophonic and trained on Western datasets, with minimal attention to polyphonic Persian music. To this end, we curated a dataset of isolated recordings of seven traditional Persian instruments and two commonly used non-traditional instruments and vocals. We then applied polyphonic augmentation (random mixing) to simulate real-world musical textures \citep{accentrecog}. Leveraging the Music undERstanding model with large-scale self-supervised Training (MERT) \citep{li2024mertacousticmusicunderstanding}, we demonstrate state-of-the-art multi-instrument classification accuracy in polyphonic Persian settings.

\textbf{The main contributions of this work are as follows:}
(1) A new publicly available dataset is introduced, consisting of isolated recordings from seven traditional Persian instruments, supplemented with violin, piano, and vocal samples.
(2) A culturally informed data augmentation strategy is proposed, designed to generate realistic polyphonic mixtures from monophonic samples through alignment of both modal structure \textit{dastgāh} and tempo.
(3) Out-of-distribution generalization is evaluated by assessing model performance on authentic Persian music recordings containing naturally occurring and complex instrument combinations.

\section{Related Work}
Early musical instrument classification relied on hand-crafted features (e.g., MFCCs, spectral descriptors) with classical models such as SVMs and KNNs, later surpassed by CNN/RNN approaches on Western-focused datasets (e.g., IRMAS, MedleyDB) \cite{1021072,859069,musicalinstrumentidentification,irmas,medleydb}. Recent progress is driven by self-supervised and foundation models that learn robust audio representations from unlabeled music, notably OpenL3, CLMR, and MERT \cite{OpenL3,CLMR,li2024mertacousticmusicunderstanding}. These models report strong results for multi-instrument tagging and are used as universal feature backbones across MIR tasks.

Despite these advances, coverage remains skewed toward Western instruments and monophonic settings, limiting generalization to polyphonic mixtures and non-Western repertoires. For Persian music, prior work addressed Radif/Dastgāh recognition and limited instrument identification from solo recordings \cite{Abbasi2014,iranianmodalmusicdastgah,BabaAli2020NavaAP}. A recent study compared self-supervised models on the Nava dataset and found MERT superior for Persian MIR tasks but only in single-label settings and without public artifacts \cite{babaali2023preprint}. Newly released resources begin to close this gap: HamNava provides multi-label, crowd-sourced annotations of Iranian classical music (instruments and vocals) and establishes baselines for cross-cultural evaluation \cite{hamnava2025}. Beyond Persian music, fresh datasets and methods also target non-Western traditions and long-form polyphonic recognition, including hierarchical detection for minority/rare instruments and song-level aggregation of snippet predictions \cite{sechet2024minority,carvalho2024,konduri2024}.

Data augmentation remains central when labeled data are scarce. Alongside time/pitch transforms and mix-based strategies for polyphony \cite{Kratimenos_2021}, recent works emphasize domain-aware mixing and label smoothing/soft-labeling for ambiguous overlaps, which are especially relevant in Persian ensembles \cite{hamnava2025}. Building on these trends, we adapt MERT for multi-label classification of Persian instruments and use polyphonic augmentation tailored to common Persian instrumentations.

\begin{figure}[t]
    \centering
    \includegraphics[width=1.0\textwidth]{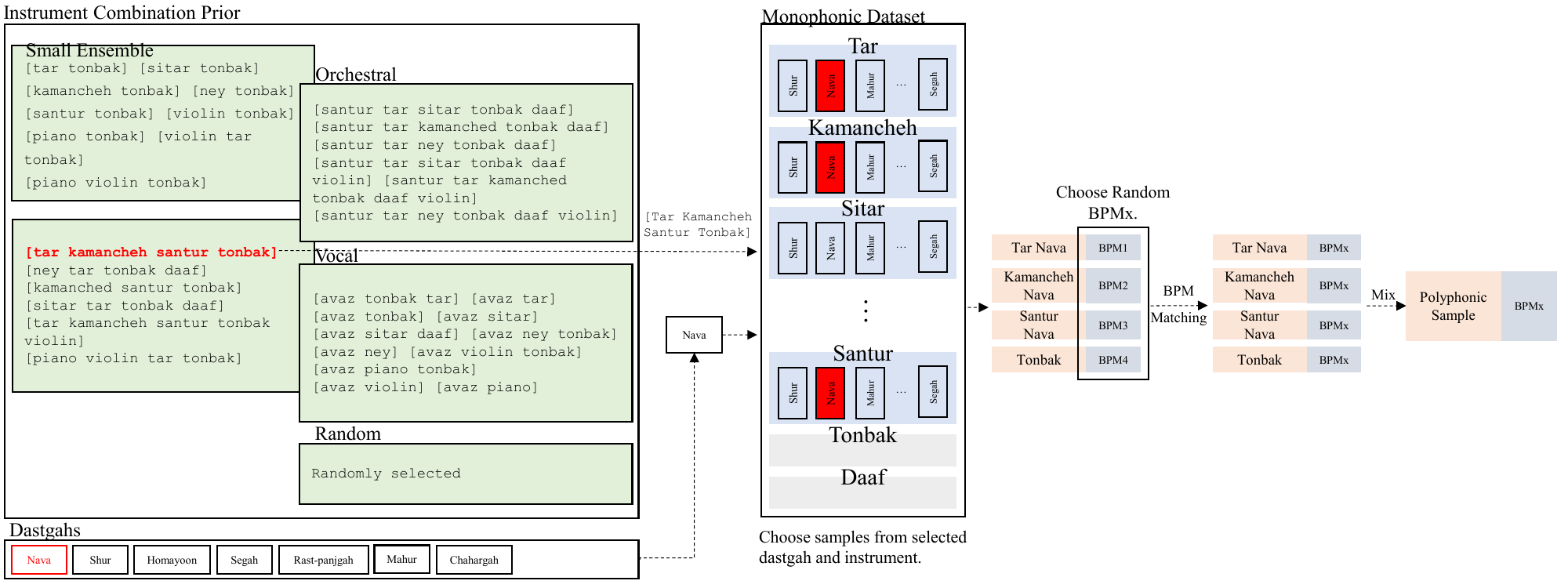}
    \caption{Dataset construction pipeline - more details at Appendix A.}
    \label{fig:dataset}
\end{figure}

\section{Method}
This research proposes a culturally informed framework for Persian musical instrument classification that bridges the gap between monophonic data scarcity and polyphonic real-world music. Starting from isolated recordings, we construct polyphonic mixtures by combining clips that share the same dastgāh and tempo, ensuring tonal and rhythmic coherence. These augmented samples emulate authentic ensemble textures and improve generalization. The resulting dataset is then used to fine-tune the MERT model with a multi-label classification head for instrument recognition in complex polyphonic contexts.
The following subsections detail the dataset preparation, augmentation strategy, and training setup of the proposed method.
\subsection{Data Preparation}
\subsubsection{Dataset Construction}
Our dataset is based on a monophonic dataset consisting of 5 second solo performances of ten diffrent instruments- Ney, Tar, Santur, Kamancheh, Daaf, Tonbak, Piano, Violin, Sitar and Avaz (Vocal - with focus on male vocals in persian traditional style). Non percussion instruments in this dataset are labeled with their key in persian music system (dastgah).
These audio clips, originating in open-source performances of various artists, include a broad spectrum of timbral and
technical characteristics as well as diverse acoustic settings.
Capturing natural
variability in audio recordings is crucial for MIR tasks, as it
improves the generalization of models to real-world conditions \cite{ivectors}. The inclusion of vocal tracks (avaz) is particularly significant, as Persian music often features highly stylized vocal
performances that are central to its cultural context.
For the construction of the dataset we used solo performances and split the tracks
from traditional Persian pieces. 
To ensure uniformity across the dataset, we removed silent
sections from each track before splitting them into 5-second
segments. This approach ensured consistency across the samples, making the dataset suitable for various machine-learning
tasks. The final dataset comprises 9 classes of instruments
plus vocal for a total of nearly 16790 audio clips, categorized in terms of key. The exact
number of samples per class is detailed in Table ~\ref{tab:monophonic}.

\begin{table}[t]
    \centering
    \caption{Comparison of constructed Polyphonic (P) samples and Monophonic (M) samples per instrument. Columns represent dastgah. Dastgah (key) is not defined for percussive instruments (Tonbak and Daaf).}
    \label{tab:monophonic}
    \centering
    \resizebox{\textwidth}{!}{%
    \begin{tabular}{lcccccccccccccccc}
        \toprule
        {} & \multicolumn{2}{c}{Chahargah} & \multicolumn{2}{c}{Homayoon} & \multicolumn{2}{c}{Mahur} & \multicolumn{2}{c}{Nava} & \multicolumn{2}{c}{Rast-Panjgah} & \multicolumn{2}{c}{Segah} & \multicolumn{2}{c}{Shur} & \multicolumn{2}{c}{Total} \\
        \midrule
         & P & M & P & M & P & M & P & M & P & M & P & M & P & M & P & M \\   
        Avaz & 2326 & 396 & 2365 & 62 & 2287 & 75 & 2357 & 125 & 0 & 0 & 2341 & 102 & 2304 & 180 & 13980 & 940 \\
        Daaf & - & - & - & - & - & - & - & - & - & - & - & - & - & - & 1 & 1302 \\
        Kaman & 1423 & 248 & 1504 & 339 & 1493 & 245 & 1416 & 260 & 1418 & 148 & 1453 & 199 & 1445 & 254 & 10152 & 1693 \\
        Ney & 1945 & 365 & 1971 & 165 & 1990 & 166 & 1914 & 130 & 1885 & 306 & 1951 & 390 & 1976 & 256 & 13632 & 1778 \\
        Piano & 1665 & 299 & 1790 & 449 & 1706 & 151 & 1645 & 97 & 0 & 0 & 1775 & 266 & 1664 & 492 & 10245 & 1754 \\
        Santur & 3017 & 240 & 3179 & 323 & 3064 & 378 & 3104 & 422 & 3089 & 269 & 3089 & 284 & 3040 & 312 & 21582 & 2228 \\
        Sitar & 1942 & 184 & 2040 & 213 & 1961 & 157 & 1928 & 82 & 1942 & 302 & 2011 & 218 & 1910 & 355 & 13734 & 1511 \\
        Tar & 4041 & 319 & 4222 & 205 & 4054 & 305 & 4168 & 477 & 4031 & 242 & 3961 & 63 & 4048 & 283 & 28525 & 1894 \\
        Tonbak & - & - & - & - & - & - & - & - & - & - & - & - & - & - & 1 & 1572 \\
        Violin & 2780 & 360 & 2915 & 253 & 2786 & 126 & 2965 & 412 & 2600 & 163 & 2887 & 642 & 2836 & 169 & 19769 & 2125 \\
        \midrule
        Total & 19139 & 2411 & 19986 & 2009 & 19341 & 1603 & 19497 & 2005 & 14965 & 1430 & 19468 & 2164 & 19223 & 2301 & 131621 & 16797 \\
        \bottomrule
    \end{tabular}
    }
\end{table}

\subsubsection{Data Augmentation Strategy}
Given the relatively small size of the initial dataset, we employed a systematic data augmentation strategy, specifically tailored to Persian music. To simulate realistic polyphonic Persian music, we developed a method in which multiple monophonic clips are combined based on a shared \textit{dastgah} and a similar tempo (BPM). This approach ensures that the synthesized tracks closely mimic authentic performances, where instruments follow both the same modal structure and a coherent rhythmic pace. For each augmented sample, a target dastgah is first selected, and all non-percussion clips are drawn from that dastgah. The BPM of the candidate clips is then estimated using librosa, one value is chosen as the base tempo, and the remaining clips are time-stretched to match it. Finally, the selected clips are mixed to create polyphonic audio containing multiple instruments and, in some cases, vocals.  

In addition to this main augmentation strategy, which combines both dastgah and tempo constraints, two alternative datasets were generated to evaluate the contribution of each factor independently. In the \textit{dastgah-only} configuration, clips share the same dastgah but are not constrained to a common tempo. Conversely, in the \textit{tempo-only} configuration, clips share the same tempo but may belong to different dastgahs. The final main dataset contains approximately 50{,}000 synthesized polyphonic samples, created predominantly using the combined dastgah and tempo approach. Figure ~\ref{fig:dataset} demonstrates the dataset construction pipeline.

\subsection{Training}
We employed the MERT-v1-330M model to classify Persian and Western musical instruments. The model extracts rich musical representations from audio inputs using multi-layer features. For multi-label classification, the original architecture was adapted by adding a classification head on top of the MERT embeddings, consisting of a fully connected layer that outputs probabilities for each instrument class. 
Instead of relying solely on the final layer, the model utilized hidden states from all layers, combining them via a learnable weighted aggregation mechanism. This dynamic approach allows the model to determine which layers contribute most to classification, improving multi-instrument recognition by leveraging both early and late-layer features.

For final classification, a two-layer neural network was used: The first layer aggregates MERT representation and the second layer outputs logits for the 10 instrument classes, with a sigmoid activation applied to produce probabilities, since multiple instruments can coexist in a sample.

For an input sample, let \(y_i \in \{0, 1\}\) represent the
true label for the i-th class, and $\hat{y_i}$ be the predicted logit for
the same class. The loss for a single sample is given by:
\begin{equation}
L = -\frac{1}{C} \sum_{i=1}^{C} 
\left[
y_i \cdot \log{\left( \sigma(\hat{y_i}) \right)} 
+ (1 - y_i) \cdot \log{\left( 1 - \sigma(\hat{y_i}) \right)}
\right]
\label{eq:loss}
\end{equation}
where C is the number of classes and $\sigma$ is the sigmoid function applied to the predicted logit.

\subsection{Experimental Setup}
For training, we fine-tuned MERT-v1-330M model on our custom dataset of Persian and Western musical instruments.  
The dataset contains approximately 50,000 5-second audio samples spanning 10 instrument classes, along with vocals. To evaluate performance in both simple and complex musical contexts, we used polyphonic samples.  
In the multi-label polyphonic setting, 50,000 samples were used for training, 108 samples for evaluation. To assess real-world generalization, we compiled a test set of 491 five-second excerpts from authentic, unedited Persian music recordings containing naturally occurring and often complex instrument combinations. All labels in this set were manually verified by two independent annotators. All models were trained with a batch size of 16 for 10 epochs, using Binary Cross-Entropy with Logits Loss (as described in Section~3), the AdamW optimizer, and a learning rate of \(1 \times 10^{-4}\).  
Model performance was assessed using ROC-AUC, Accuracy, and F1-score metrics.

\subsection{Results}
Table~\ref{tab:results} reports the performance of the three augmentation strategies, along with a random-mixing baseline, on the Persian instrument recognition task. Among the evaluated methods, the \textit{Dastgah-only} configuration achieved the highest accuracy (0.841) and F1-score (0.669) when tested on real-world polyphonic recordings. In contrast, the \textit{Dastgah+BPM} strategy yielded the highest ROC-AUC (0.795), indicating superior ranking capability across label thresholds. The \textit{BPM-only} variant underperformed compared to the other augmentation schemes, suggesting that tonal coherence (shared dastgah) has a stronger impact on recognition than rhythmic alignment alone.  
Overall, these results demonstrate that constraining augmented samples to share a common dastgah is a key factor for improving generalization to authentic music, while the addition of tempo alignment offers further benefits in ROC-AUC, highlighting complementary effects between tonal and temporal consistency. Further analyses are presented in Appendix C.

\begin{table}[t]
    \centering
    \caption{Performance comparison of different augmentation strategies on the Persian instrument recognition task, evaluated on the test set comprising authentic Persian music recordings.}
    \label{tab:results}
    \begin{tabular}{lccc}
        \toprule
        \textbf{Data Augmentation} & \textbf{Accuracy} & \textbf{ROC-AUC} & \textbf{F1-score} \\
        \midrule
        Random & 0.794 & 0.750 & 0.606 \\
        
        BPM & 0.807 & 0.764 & 0.617 \\
        Dastgah & \textbf{0.841} & 0.780 & \textbf{0.669} \\
        Dastgah+BPM & 0.823 & \textbf{0.795} & 0.652 \\
        \bottomrule
    \end{tabular}
\end{table}

\section{Conclusion}
We presented a Persian instrument classification framework built on the MERT architecture and trained using a newly constructed dataset enriched through culturally informed polyphonic data augmentation. By systematically evaluating three augmentation strategies, shared dastgah, shared tempo, and the combination of both, we showed that dastgah is the dominant factor for improving accuracy and F1, while the combination of dastgah and tempo alignment yields the highest ROC-AUC.  
This study highlights the importance of culturally grounded augmentation methods for enhancing model robustness in low-resource MIR scenarios. The proposed dataset and augmentation techniques have potential applications in automatic music tagging, and data-efficient music generation.

\bibliographystyle{plainnat}  
\bibliography{main}

\newpage
\appendix
\section*{Appendix A: Dataset Construction Procedure}

Each dataset sample was constructed according to the following procedure:

\begin{enumerate}
    \item \textbf{Selection of Dastgāh:} \\
    A base modal system (\emph{dastgāh}) was chosen at random from the seven canonical Persian music systems: 
    \emph{Nava}, \emph{Shur}, \emph{Homayoon}, \emph{Segāh}, \emph{Rāst-Panjgāh}, \emph{Māhur}, and \emph{Chahārgāh}.

    \item \textbf{Partitioning by Ensemble Prior:} \\
    The dataset was divided into five equally sized partitions, each corresponding to a distinct ensemble prior:
    \begin{itemize}
        \item small ensembles,
        \item medium ensembles,
        \item orchestral tracks,
        \item tracks with vocals,
        \item random instrument combinations.
    \end{itemize}
    Each partition contained the same number of samples, ensuring balanced representation across ensemble types.

    \item \textbf{Instrument Selection and Sampling:} \\
    Within each partition, a specific instrument combination was selected according to its prior. For every instrument in the chosen combination, an audio sample was drawn in the specified \emph{dastgāh}.

    \item \textbf{Tempo Alignment:} \\
    The beats per minute (BPM) of all selected samples were computed. One of these BPM values was designated as the baseline tempo. All other samples were time-stretched to match this baseline, thereby standardizing tempo across the ensemble.

    \item \textbf{Polyphonic Mixing:} \\
    The tempo-aligned samples were subsequently mixed, producing a polyphonic audio sample representative of the given \emph{dastgāh} and ensemble prior.
\end{enumerate}

This process ensured systematic variation along two key axes: 
(i) \emph{modal diversity} (through the seven \emph{dastgāhs}) and 
(ii) \emph{ensemble texture} (through balanced priors). 
The resulting dataset exhibits both stylistic authenticity and structural balance, making it suitable for downstream tasks in computational ethnomusicology and generative modeling.

\appendix
\section*{Appendix B: Beat Tracking with \texttt{librosa.beat.beat\_track}}

To estimate tempo and beat positions, we employed the 
\texttt{librosa.beat.beat\_track} function, a widely used algorithm for beat 
tracking in music information retrieval. The method combines onset detection, 
tempo estimation, and dynamic programming to provide both the global tempo (in 
beats per minute) and the sequence of beat times.

\textbf{Algorithm overview.} 
First, the audio signal is transformed into a spectral representation, from 
which an \emph{onset strength envelope} is derived. This one-dimensional curve 
highlights moments of increased spectral energy corresponding to note or drum 
onsets. Next, the periodicity of the envelope is analyzed to estimate the most 
likely tempo. The algorithm applies a bias towards musically plausible tempos, 
while accounting for common ambiguities such as half- and double-tempo errors. 
Finally, beat positions are selected using a \emph{dynamic programming 
procedure} that balances two criteria: (i) alignment with strong onsets and 
(ii) regularity consistent with the estimated tempo. This produces an optimal 
sequence of beat times.

\textbf{Motivation for use.} 
This approach is computationally efficient, generalizes well across a wide 
range of musical genres, and requires no training data. It provides a robust 
foundation for higher-level tasks such as feature synchronization, music 
segmentation, remixing, or time-aligned annotation.

\section*{Appendix C: Experiments}
\label{appendix:appendixC}
To evaluate the trained models more precisely and analyze the recognition performance of individual instruments, Table~\ref{tab:per-label-accuracy-four} reports the per-label accuracy for all ten instrument classes. This detailed breakdown provides a clear view of how each model performs across both Persian (non-Western) and Western instruments. Although Western instruments such as \emph{piano} and \emph{violin} generally achieve slightly higher detection accuracy, the results indicate that Persian instruments like \emph{kamanche}, \emph{daaf}, and \emph{santur} are also recognized with strong reliability. Overall, the table demonstrates that the proposed models achieve balanced performance across culturally diverse instrument categories.

\begin{table}[h]
\centering
\caption{Per-label accuracy for four models in the multi-label setting.}
\label{tab:per-label-accuracy-four}
\small
\begin{tabular}{lcccc}
\toprule
\textbf{Label} & \textbf{Dastgah+BPM} & \textbf{Dastgah} & \textbf{BPM} & \textbf{Random} \\
\midrule
kaman & 0.866 & \textbf{0.882} & 0.849 & 0.835 \\
ney      & 0.776 & 0.788 & \textbf{0.794} & 0.752 \\
santur   & \textbf{0.859} & 0.823 & 0.794 & 0.823 \\
sitar    & 0.711 & \textbf{0.859} & 0.701 & 0.760 \\
tar      & 0.715 & 0.737 & \textbf{0.747} & 0.707 \\
tonbak   & 0.768 & \textbf{0.804} & 0.798 & 0.798 \\
daaf     & \textbf{0.870} & 0.847 & 0.813 & 0.715 \\
avaz     & 0.908 & \textbf{0.929} & 0.894 & 0.898 \\
piano    & \textbf{0.849} & 0.821 & 0.782 & 0.821 \\
violin   & 0.906 & \textbf{0.923} & 0.898 & 0.835 \\
\bottomrule
\end{tabular}
\end{table}

\begin{figure}[t]
    \centering
    \includegraphics[width=0.9\linewidth, height=0.5\linewidth]{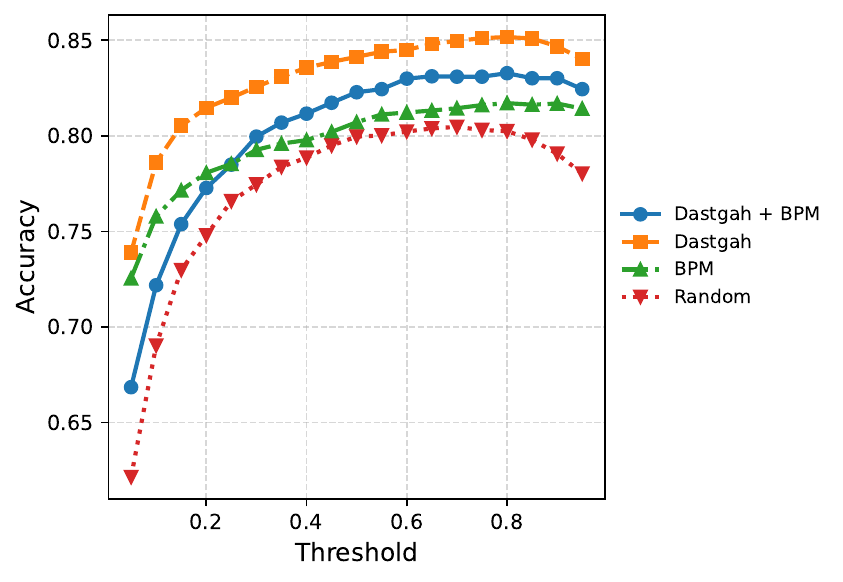}
    \caption{Accuracy of different models across varying decision thresholds.
    Each curve represents a distinct model configuration.}
    \label{fig:accuracy-threshold}
\end{figure}

Figure~\ref{fig:accuracy-threshold} illustrates how the classification accuracy of each model varies with different decision thresholds. This analysis helps determine the optimal trade-off between sensitivity and precision for multi-label prediction. As the threshold increases, accuracy generally stabilizes, revealing the robustness of the proposed models across a wide range of cutoff values.

Table~\ref{tab:ablation-mert} reports the results of the ablation study comparing two training strategies: freezing the MERT encoder (and training only the classifier head) versus applying LoRA fine-tuning with a rank of 16.
Overall, both settings achieve reasonable performance, with the LoRA configuration slightly improving over the frozen encoder in most cases—for example, the Dastgah+BPM model increases its ROC–AUC from 0.757 to 0.777 and accuracy from 0.779 to 0.795.
However, when compared to full fine-tuning of both the MERT encoder and classifier head, these partial adaptation strategies still underperform. Full fine-tuning consistently achieves the highest ROC–AUC and accuracy across all model configurations, indicating that updating the entire encoder provides better task-specific representation learning than either freezing or parameter-efficient adaptation methods.

\begin{table}[t]
\centering
\caption{Ablation study on training strategy. Comparison between freezing the MERT encoder (training only the classifier head) and fine-tuning with LoRA (rank = 16).}
\label{tab:ablation-mert}
\small
\begin{tabular}{lcccc}
\toprule
\multirow{2}{*}{\textbf{Model}} & 
\multicolumn{2}{c}{\textbf{Frozen MERT Encoder}} & 
\multicolumn{2}{c}{\textbf{LoRA (rank = 16)}} \\
\cmidrule(lr){2-3} \cmidrule(lr){4-5}
 & ROC–AUC & Accuracy & ROC–AUC & Accuracy \\
\midrule
Dastgah + BPM & 0.757 & 0.779 & \textbf{0.777} & \textbf{0.795} \\
Dastgah       & 0.754 & 0.733 & 0.757 & 0.713 \\
BPM           & 0.768 & 0.782 & 0.776 & 0.749 \\
Random        & 0.772 & 0.786 & 0.770 & 0.773 \\
\bottomrule
\end{tabular}
\end{table}

\end{document}